\documentclass[a4paper,12pt,epsf]{article}

\usepackage{t1enc}
\usepackage[latin1]{inputenc}
\usepackage[final]{graphicx}
\usepackage{amsfonts}

\newcommand{\bin}[2]{{ #1 \choose #2 }}
\newcommand{\var}{\gamma}
\newcommand{\moy}{\alpha}
\newcommand{\mona}{\alpha}
\newcommand{\binf}{\beta}
\newcommand{\bsup}{\beta}
\newcommand{\mont}{\Upsilon}
\newcommand{\vev}[1]{\left<  #1 \right>}

\title{\LARGE\bf Asymmetric evolving random networks}
\date{}
\author{}

\begin{document}

\maketitle

\vspace{-1.2cm}

\centerline{\large St\'ephane Coulomb\footnote[1]{Email:
    coulomb@spht.saclay.cea.fr} and  Michel Bauer\footnote[2]{Email:
    bauer@spht.saclay.cea.fr}} 

\vspace{.3cm}

\centerline{\large Service de Physique Th\'eorique de
  Saclay\footnote[3]{\it Laboratoire de la Direction des Sciences de
    la Mati\`ere du Commisariat \`a l'Energie Atomique, URA2306 du CNRS}}

\vspace{.3cm}

\centerline{\large CE Saclay, 91191 Gif sur Yvette, France}

\vspace{.3cm}

\begin{abstract} We generalize the poissonian evolving random graph
  model of \cite{bb} to deal with arbitrary degree distributions. The
  motivation comes from biological networks, which are well-known to
  exhibit non poissonian degree distribution. A node is added at each
  time step and is connected to the rest of the graph by oriented
  edges emerging from older nodes. This leads to a statistical
  asymmetry between incoming and outgoing edges. The law for the
  number of new edges at each time step is fixed but arbitrary.
  Thermodynamical behavior is expected when this law has a large time
  limit. Although (by construction) the incoming degree distributions
  depend on this law, this is not the case for most qualitative
  features concerning the size distribution of connected components, as
  long as the law has a finite variance. As the variance grows above
  $1/4$, the average being $<1/2$, 
  a giant component emerges, which connects a finite
  fraction of the vertices. Below this threshold, the distribution of
  component sizes decreases algebraically with a continuously varying
  exponent. The transition is of infinite order, in sharp contrast
  with the case of static graphs. The local-in-time profiles for the
  components of finite size allow to give a refined description of the
  system.
\end{abstract}

\section{Introduction}

Evolving graphs arise naturally in the modelization of communication
networks, but also of social organizations and biological phenomena :
brain formation, genetic regulations, etc. Until recently, quantitative
data were scarce, but the situation is changing very quickly
\cite{barab,kepes,lee}. In many important cases the laws governing the
evolution of the network are unknown but non deterministic, and the
final number of nodes is rather large.  This explains why physicists
(see the reviews \cite{barab,doro1} and references therein) have been
developping recently random graph techniques in the thermodynamic
limit to understand some peculiar features, the most salient being
large degree distributions, that seem to occur in an unexpected
variety of situations.

Many cases of interest lead to oriented graph models, for which the in-
and out-degree distributions can be governed by different laws. This
was one of the basic observations in the study of the yeast genetic
regulatory network presented in \cite{kepes}: a single gene may
participate to the regulation of many other genes -- the law for
out-degrees seems to be large --, but each gene is only regulated by
a few other genes -- the law for in-degrees seems to have finite
moments. A biological interpretation for the asymmetry is that the few
promoter-repressor sites for each gene bind only to specific proteins,
but that along the genome many promoter-repressor sites are
homologous. One of our aims is to work with a model that incorporates
such an asymmetry. 

We shall follow quite closely the philosophy of \cite{bb}, and in
particular pay attention not only to global quantities, but also to
local-in-time profiles. We feel that for evolving networks this is a
crucial condition to extract relevant information, because as we shall
see, global quantities (averaged over time) give a distorded view of
the network. On the other hand, it is hard experimentally to access local quantities, either because the ages are not known, or because
their consideration would reduce the statistics down to an
unacceptable level. But the amount of available information is growing
very rapidly, and one can hope that local
quantities will become accessible in a near future.

\vspace{.3cm}

The model we study is the natural evolving cousin of the static
maximal entropy model with given in-degree distribution
\cite{bb2,whittle}. Starting from a single
vertex at time $1$, a new vertex is created at each time step -- so
that at time $t$, the size of the system, i.e. the number of vertices,
is $t$ -- and new oriented edges are created with specified
probabilistic rules. An arbitrary probability generating function $T$
encodes the parameters of the model in the thermodynamic (large $t$)
limit.  Precise definitions are given in the next section. 

Our main results are the following~:

The global and local in-degree distributions are given by $T$, see eq.(\ref{incoming}), whereas
the global out-degree distribution is geometric with average $T'(1)$,
see eq.(\ref{outgoing}).
The local out-degree distribution is poissonian but age dependant, eq.(\ref{outgoinglocal}).
In the case of a static maximal entropy random graph with the same
in-degree distribution (\cite{bb2,whittle}), the distribution of
out-degrees would be poissonian.

The global structure of connected components is studied via a
generating function which satisfies a differential equation, leading
to recursion relations for the distribution of component sizes, see 
eq.(\ref{evol_2}). This is illustrated by analytical computations in
the case when $T(z)$ is a polynomial of degree $1$. 

The general criterion for the absence of a component
containing a finite fraction of the sites in the thermodynamic limit
is that the average $\moy \equiv T'(1)$ be $\leq 1/2$ and the variance
$\var \equiv T'(1)+T''(1)-T'(1)^2$ be $\leq 1/4$. In that case the system
contains components whose sizes scale like a power of the total size
of the graph, see eq.(\ref{grosclu}); equivalently, the probability
distribution for component sizes has an algebraic queue, see
eq.(\ref{kdecrbelow}). Above the threshold (when $T'(1) > 1/2$ or
$\var >1/4$) this probability distribution
is defective but decreases exponentially see eq.(\ref{kdecrabove}).
The boundary separating the percolating and non
percolating phases is $\var =1/4$ and $\moy < 1/2$. In the
percolating phase, but close to this boundary, $\var-1/4$ is $>0$ but
small and the giant
component is exponentially small, see eq.(\ref{sing}).  This
situation, somehow reminiscent of the Kosterlitz-Thouless transition,
had already been observed in a variety of models \cite{new,doro2,bb}.

We compare the percolation criterion with the one that emerges from
the study of a static maximal entropy random graph with the same
in- and out-degree distributions as our evolving graph
(\cite{bb2,whittle}). In the static case, the growth of the giant
component is generically linear close to the threshold. But we show
that if $T(z)$ leads to a percolating static graph, it is
automatically percolating for the evolving graph model. The intuitive
explanation lies in the inhomogeneities of the evolving graph : the
environment of an old vertex is denser than what a static model
produces. And indeed, the giant component profile close to threshold
is very asymmetric, see eqs.(\ref{rhoinf_1},\ref{rhoinf_0}).  

In the thermodynamic limit, the finite components are trees, and we
derive a direct enumeration formula to count for their abundance
eq.(\ref{arbres_2}).  This can be used to describe all local in time
profiles of finite components, a result we also recover for one time
quantities via a generating function aproach eq.(\ref{evol_3}). In the
appendix, we give a proof of the equivalence of the generating
function approach and tree enumeration, a question that was left
opened in \cite{bb}.

We have confronted our analytical results with numerical simulations
whenever possible, our preferred  example being when $T$ is a
geometric distribution.

\vspace{.5cm}

\noindent Aknowledgement : we thank Denis Bernard for a careful
reading of the manuscript and for his kind permission to include the
results of section.~\ref{simpleex}, that where obtained with him some time ago.

\section{The model}

 The random graphs we consider are constructed according to the
following rules :
\begin{enumerate}
\item For $t=1,2,\cdots $, we denote by ${\mathcal G}_t$ the set of
  simple graphs with vertex set $V_t=[1,\cdots ,t]$, i.e. the set of
  pairs $(V_t,E_t)$ where $E_t$, the set of edges of the graph, is a
  subset of $\{(i,j),1 \leq i<j \leq t\}$. We orient the graph
  $(V_t,E_t)$ by saying that $(i,j)\in E_t$ is an edge \textit{from} $i$ \textit{to}
  $j$~: an edge always goes from an older to a younger
  vertex. Note that ${\mathcal G}_1=\{(\{1\},\emptyset)\}$ contains a
  single graph, made of one vertex but no edge.
\item An evolving graph is a sequence $G=\left( G_t \right)_{t \geq
    1}=\left((V_t,E_t) \right)_{t \geq 1}$, 
  where $G_t \in {\mathcal G}_t$ for $t \geq 1$, and for $t \geq 2$,
  $G_{t-1}$ is the induced subgraph of $G_t$ obtained by removing
  vertex $t$ and all edges adjacent to it. Stated differently, $G_t$
  for $t\geq 2$ is obtained from $G_{t-1}$ by adding vertex $t$ and
  some edges of the form $(i,t)$, $i<t$. This model of evolution
  implies that knowing $G_t$ is equivalent to knowing
  $G_1,\cdots ,G_{t-1},G_t$. We write ${\mathcal G}$ for the set of
  evolving graphs. 
\item In addition to these rules of construction, we put a probability
  measure $p$ on the set of evolving graphs. If $G$ is an evolving
  graph, we denote by $\hat{G}_t$ the number of edges arriving at
  vertex $t$ in $G_t$ (which is the same as the number of edges
  arriving at $t$ in $G_{t'}$ for any $t'\geq t$). We want to fix the
  probability distributions for the number of edges arriving at
  vertices $1,2,\cdots $, i.e. we impose $p(\{G \in {\mathcal
    G},\hat{G}_t=k \})=\tau_{t,k}$ for a given sequence of probability
  distributions $\{\tau_{t,k}, t\geq 1, 0 \leq k\leq
  t-1\}$,$\tau_{t,k}\geq 0$, $\sum_k \tau_{t,k}=1$. Since this criterion
  is far from fixing unambiguously the probability law, we shall only
  be interested in the one which doesn't introduce any other bias,
  i.e. the one which maximizes entropy. Then a simple computation shows
  that for $H_t$ in ${\mathcal G}_t$, $$p(\{G \in {\mathcal G},
  G_t=H_t\})=\prod _{1 \leq s \leq t}
  \frac{\tau_{s,\hat{H}_{s}}}{\bin{s-1}{\hat{H}_{s}}}.$$
\end{enumerate}

Note that we did not worry about the size of the sequence defining an
evolving graph. In fact, we shall as often as possible suppose it is
infinite.

\section{Notations and basic results}

\subsection{Generating functions}

For any vertex $v \geq 1$, let $T_v$ be the generating function for
the incoming degree distribution :
$$T_v (z) \equiv \sum_{k=0}^{v-1} \tau_{v,k} z^k $$
\\
We shall be mostly interested in the thermodynamic limit, i.e. in $t$
independant features of the large $t$ behavior of $G_t$ when $G$ is a
random element in $({\mathcal G},p)$. A typical example is the
behavior of $\lim_{ t\rightarrow \infty}|E_t|/|V_t|$ \footnote{Here
  and in the sequel, $|S|$ denote the number of elements of the finite
  set $S$.}.  Obviously, this limit does not exist for any \textit{a
  priori} choice of the $\tau_{v,k}$'s. The study of thermodynamic
convergence for this quantity and others would be of independent
interest, but we shall see that a simple assumption ensures that many
quantities of interest have a thermodynamic limit : we impose that the
sequence of functions $T_v (z)$ converges to a probability generating
function
\begin{equation}
T (z) \equiv \sum_{k \geq 0} {\tau_{k} \over k!}  z^k
\end{equation}
in such a way that the sequence of averages and variances converges to
that of $T (z)$.

This property implies in particular that, for any $k$, $\tau_{v,k}$
tends to ${\tau_{k} \over k!}$ for large $v$. Moreover, we will often
use the fact that $T(1)=1$. We use the notations $\moy \equiv T'(1)$
and $\var \equiv T''(1)+T'(1)-T'(1)^2$ for the average and variance of
$T$.

As a technical hypothesis, to avoid
several pathologies, we shall always implicitely assume that
$\tau_0=T(0)>0$.

\subsection{A preliminary formula}

We shall sometimes be lead to consider situations in which a new
vertex $t+1$ appears and connects to the rest of the graph avoiding $m$
forbidden vertices. This will happen, for instance, if one is
interested in the distribution of connected components in the graph.
In this section we would like to give a general formula for such
situations. Hence, let $\mona_{t+1,m}$ be the probability for a new
vertex $t+1$ not to connect to $m$ given vertices. This probability is
given by the formula :
$$\mona_{t+1,m} \equiv \sum_{n \geq 0} \tau_{t+1,n} {{t-m \choose n} \over
  {t \choose n}}$$
With $\mona_{t+1,m}(z) \equiv \sum_{n \geq 0} \tau_{t+1,n} z^n {{t-m
    \choose n} \over {t \choose n}}$, 
we notice that  $\mona_{t+1,m+1}(z)=\mona_{t+1,m}(z)-{z \over t-m}
\mona'_{t+1,m}(z)$. Since $\mona_{t+1,0}(z)=T_{t+1}(z)$, we have :
\begin{eqnarray*}
\mona_{t+1,m}(z) & = & T_{t+1}(z)-{{m \choose 1} z \over t} T_{t+1}'(z) \\
 & & \ +{{m \choose 2} z^2 \over t(t-1)} T_{t+1}''(z)-\cdots (-)^m {{m \choose m} z^m \over
  t\cdots (t-m+1)} T_{t+1}^{(m)}(1) \\ \mona_{t+1,m} & = & 1-{{m \choose 1}
  \over t} T_{t+1}'(1) \\
 & & \ +{{m \choose 2} \over t(t-1)} 
T_{t+1}''(1)-\cdots (-)^m{{m \choose m} \over t(t-1)\cdots (t-m+1)} T_{t+1}^{(m)}(1)
\end{eqnarray*}

In the thermodynamic limit, we shall only need the large $t$ finite
$m$ approximation
\begin{equation}
\mona_{t+1,m} \simeq 1-{\moy  m \over t}.
\label{theta}
\end{equation}

These simple formul\ae\ may now be used to calculate a few quantities.
We'll first describe the degree distributions of incoming and outgoing
edges, whereas the two following sections will present the main
relations governing the distributions of connected components of the
graphs.

\section{Degree distribution}

\subsection{Notations}
Let 
\begin{enumerate}
\item $l_j^-(t)$ (resp. $l_j^+(t)$) be the number of incoming
  (resp. outgoing) edges at a vertex $j$ at time $t$
\item $v_k^-(t)$ (resp. $v_k^+(t)$) be the number of vertices with $k$ 
  incoming (resp. outgoing) edges at time $t$.
\end{enumerate}
With these notations, the edge distributions are described by the
generating functions :
$$\nu^\pm_t(z) \equiv {1 \over t} \sum_{0 \leq k < t} \vev{
  v_k^\pm (t) } z^k = {1 \over t} \sum_{1 \leq j \leq t} \vev{
  z^{l_j^\pm (t)} }$$
 In the thermodynamic limit, it is possible to give an expression for
 these generating functions.

\subsection{In-degree distribution}
Up to a normalisation factor, the probability for vertex $j$ to have
$k$ incoming edges is $\tau_{j,k}$.This shows that 
$$\vev{ z^{l_j^- (t)} } = \sum_{k \geq 0} \tau_{j,k} z^k$$
In the thermodynamic limit, this becomes
\begin{equation}
\label{incoming}
\vev{ z^{l_j^- (t)} } = T(z)
\end{equation}
Hence, in the thermodynamic limit, the probability for any vertex to
have $k$ incoming edges is given by ${\tau_k \over k!}$, which is
correctly normalized.
\\
This implies in particular that, in the thermodynamic limit, the
average number of edges coming to a vertex is precisely $\moy $.

\subsection{Out-degree distribution}
Any edge emerging from a vertex $j$ can be seen as an edge arriving at a
vertex younger than $j$. Let  $t>j$ be the time at
which one observes the number of edges emerging from $j$. We define
$\sigma\equiv j/t$. The quantity $\vev{ z^{l_{\sigma t} ^+(t)} }$ is
given by the formula :
$$\sum_k z^k \sum_{\sigma t < j_1<\cdots <j_k \leq t} \prod_{1 \leq i \leq
  k} \left( \sum_{0 \leq k' \leq j_i-2} \tau_{j_i,k'+1} {{j_i-2
      \choose k'} \over {j_i-1 \choose k'+1}} \right) \prod_{j < j'
  \leq t ; j' \neq j_i } \left( \sum_{0 \leq k' \leq j'-2 }
  \tau_{j',k'} {{j'-2 \choose k'} \over {j'-1 \choose k'}} \right)$$
Indeed, the probability for $j$ to have $k$ outgoing edges is obtained
by summing over $j_1,\cdots ,j_k>j$ the probabilities for $j$ to be linked
to these $k$ vertices and to no other vertex. Each of this
probabilities can easily be calculated from the following relations :
\begin{eqnarray*}
\sum_{0 \leq k' \leq j_i-2} {\tau_{j_i,k'+1} \over {j_i-1 \choose
    k'+1}} {j_i-2 \choose k'} & = 
& {T_{j_i}'(1) \over j_i-1} \\
\sum_{0 \leq k' \leq j'-2 } {\tau_{j',k'} \over {j'-1 \choose k'}}
{j'-2 \choose k'}& = 
& 1- {T_{j'}'(1) \over j'-1 }
\end{eqnarray*}
For fixed $\sigma$ and large $t$, the second product simplifies to
$\sigma^{\moy }$ and the average we look for becomes :
\begin{eqnarray*}
\vev{ z^{l_{\sigma t} ^+(t)} } & \simeq & \sigma^{\moy } 
\sum_k z^k \sum_{ \sigma t < j_1<\cdots <j_k \leq t} {\moy ^k 
\over j_1\cdots j_k} \\
 & \simeq &  \sigma^{\moy } \sum_k z^k {1 \over k!} 
\left[ \ln t - \ln \sigma t \right]^k \moy ^k \\
\end{eqnarray*}

Hence, for fixed $\sigma$, 
\begin{equation}
\label{outgoinglocal}
\vev{ z^{l_{\sigma t} ^+(t)} } \rightarrow  e^{-(z-1) \moy  \ln
  \sigma}
\end{equation}
 for large $t$, so the local out-degree distribution is
Poissonian with $\sigma$-dependent parameter $\moy  \ln
  \sigma$. Integrating $\sigma$ between $0$ and $1$ yields the  asymptotic
value  for the generating function $\nu_t^+ (z)$ :
\begin{equation}
\label{outgoing}
\nu_t^+ (z)\rightarrow  {1 \over 1+ \moy  (1-z)}
\end{equation}
at large $t$. 

Identifying the term of degree $k$ in $z$ in the development of this
function finally yields the probability for a vertex to have $k$
outgoing edges :
$$p^+(k)={\moy ^k \over \left( 1+\moy  \right) ^{k+1}}$$

We see that the distribution of outgoing edges is geometric and
depends on the probability distribution $T$ only through the average
number of incoming edges $\moy $.

\subsection{Mixed distribution}
In our model, the number of in and out edges $l_j^+(t)$ and $l_j^-(t)$
at a given vertex are
independent by construction. The generating function for the mixed
degree distribution is
$$\nu_t(z_+,z_-) \equiv {1 \over t} \sum_{k_+,k_-} \vev{
  v_{k_+,k_-}(t) } z_+^{k_+} z_-^{k_-} = {1 \over t} \sum_{1
  \leq j \leq t} \vev{ z_+^{l_j^+(t)} z_-^{l_j^-(t)} }$$
which is easily obtained from eqs.(\ref{incoming},\ref{outgoing}) :
\begin{equation}
\nu(z_+,z_-)={T(z_-) \over 1+ \moy  (1-z_+)}.
\end{equation}
The local counterpart would easily follow from
eqs.(\ref{incoming},\ref{outgoinglocal}). 

\section{Connected components}
In this section we give formul\ae\ for the number of components of
size $k$ in the thermodynamic limit (paragraph~\ref{global}), and for
the time distribution of these components (paragraph~\ref{local}).
This derivation makes some natural assumptions of self averaging. In
paragraph \ref{trees}, we evaluate without such assumptions the
contribution of individual graphs to the weights of random graphs, and
show that trees dominate the thermodynamic limit. We use this result
in the appendix to compute directly the generating function for the
number of components of size $k$, and show that it coincides with the
one given in paragraph~(\ref{global}).

\subsection{Global-time results}
\label{global}
Let $N_k(t)$ be the number of connected components of size $k$ at time
$t$ for a given graph, and $N_t(z)$ the corresponding generating function :
$$N_t(z) \equiv \sum_{k \geq 1} N_k(t) z^k$$
To evaluate this
function, we shall investigate its evolution between times $t$ and
$t+1$. When, at time $t+1$, the new vertex is added and randomly connected to
some of the older vertices, the components to which these vertices
belonged are destructed and collapsed into a new, bigger, component.
More precisely, let us put $n_k(t)$ for the number of components of
size $k$ connected to vertex $t+1$ when it is added to the graph. For all $k$,
$n_k(t)$ components of size $k$ are thus destroyed between times $t$
and $t+1$, while a component of size $1+\sum_k k n_k(t)$ is created.
To summarize this :

\begin{equation}
N_{t+1}(z)-N_t(z)=-\sum_{k}n_k(t) z^k+z^{1+\sum_k k n_k(t)}
\label{evol_1}
\end{equation}

Now, the point is to average this equality over all
graphs of size $k$ and all possible connections of vertex $t+1$ with
these graphs. 

First of all, we shall propose a compact expression for $\vev{ z^{n_k(t)}
}$ valid for all $t$. We will deduce from it an expression of
$\vev{ n_k(t) }$ before calculating the average of $z^{1+\sum_k
  k n_k(t)}$. Finally we will give an evolution equation valid in the
thermodynamic limit.

\paragraph{Exact derivation of $\vev{ z^{n_k(t)} }$}

This average is given by 
$$
\sum_{G_t} \sum_{j \geq 0} z^j w_{t+1}(k,j),$$
where $w_{t+1}(k,j)$
is the total weight of graphs of size $t+1$ obtained from a graph
$G_t$ of size $t$ in such a way that, when it was added, vertex $t+1$
connected to $j$ components of size $k$. Hence, $w_{t+1}(k,j)$ is the
term of degree $0$ in $\omega$ of the Laurent serie :
$$w_{G_t} \sum_{n \geq 0} {\tau_{t+1,n} \over {t \choose
    n}}\omega^{-n} {N_k(t) \choose j} \left( \sum_{i=1}^k \omega^i {k
    \choose i} \right)^j \sum_{p \geq 0} \omega^p {t-k N_k(t) \choose
  p}$$
In this expression,
\begin{itemize}
\item $n$ is the number of edges connecting vertex $t+1$ to $G_t$ ;
\item $i$ is the number of edges connecting vertex $t+1$ to a given
  component of size $k$ ;
\item $p$ is the number of edges connecting vertex $t+1$ to vertices of
  $G_t$ which don't belong to a component of size $k$.
\end{itemize}
The sums over $i$, $j$ and $p$ simplify, so that $\vev{z^{n_k(t)} }$
is the term of degree $0$ in  $\omega$ of
$$\sum_{G_t,n\geq 0} w_{G_t} {\tau_{t+1,n} \over {t \choose
    n}}\omega^{-n} \left( z(1+\omega)^k+1-z
\right)^{N_k(t)}(1+\omega)^{t-kN_k(t)} $$
If we rewrite the sum over the $G_t$'s as an average over graphs of
size $t$ and expand the two parenthesis, we get
$$\vev{z^{n_k(t)} }  =  \vev{ \sum_{n,m \geq 0} {\tau_{t+1,n}
    \over {t \choose n}} {N_k(t) \choose m} \left( 1-z \right) ^m
  z^{N_k(t)-m}{t-km \choose n} }$$

In this formula, we recognize
$\mona_{t+1,m} \equiv \sum_{n \geq 0} {\tau_{t+1,n} \over {t
    \choose n}} {t-m \choose n}$, which is the probability that
vertex $t+1$ doesn't connect to $m$ given vertices. The average we
need becomes :
\begin{equation}
\vev{z^{ n_k(t)} } =\sum_{m\geq 0}\mona_{km} \vev{ {N_k(t)
    \choose m} z^{N_k(t)-m} \left( 1-z \right) ^m }  
\label{composantes_k}
\end{equation}

This formula gives in particular the average number of components of size 
$k$ connected to the new vertex :
\begin{equation}
\vev{ n_k(t) } =  \vev{  N_k(t) } \left(
  1-\mona_{t+1,k} \right)
\label{moy_simpl}
\end{equation}

\paragraph{Exact derivation of $\vev{ z^{\sum_k n_k(t)} }$}

The only difference with the computation we've just described is that the 
sum over $j$ is replaced by sums over all possible $j_k$'s, while
$z^j$ becomes $z^{\sum_k k j_k}$. Omitting the details, we just give
the result, which has the same form as formula (\ref{composantes_k}) :
$$
 \vev{ z^{\sum_{k} k n_k} }=\sum_{ \{ m \} } \mona_{t+1,\sum_k k
   m_k } \vev{ \prod_k \left( {N_k(t) \choose m_k} 
z^{k\left( N_k(t)-m_k \right) } \left( 1-z^k \right) ^{m_k} \right) }
$$
In this equation, the sum is performed over all sequences
$m_1,m_2,\cdots $.
\\
This expression can be simplified if we note that $\mona_{t+1,m}$ is a
weighted sum of Newton coefficients, so that 
$$\mona_{t+1,m}=\sum_{n \geq 0} \oint {dx \over x^{n+1}} {\tau_{t+1,n}
  \over {t \choose n}}
\left( 1+x \right) ^{t-m},$$
where the symbol $\oint$ denotes the contour integral ${1 \over
  2i\pi}\int$ along a small contour surrounding the origin.
\\
Evaluating the sum over the $m_k$'s now yields

\begin{equation}
 \vev{ z^{\sum_{k} k n_k} }=\sum_n {\tau_{t+1,n} \over {t
     \choose n}} \oint {dx \over x^{n+1}} \vev{ \prod_k \left( z^k 
\left( 1+x \right) ^k +1-z^k 
\right) ^{N_k(t)} }
\label{toutes_comp}
\end{equation}

\paragraph{Evolution equation for connected components}
Formula (\ref{evol_1}) can now be written in a simpler form using
equalities (\ref{moy_simpl}) and
(\ref{toutes_comp}). With $T_t(1)=1$, this leads to :
\begin{eqnarray}
\label{exact}
\vev{ N_{t+1}(z) }-\vev{ N_t(z) } & = & \sum_k \vev{
  N_k(t) } \left(\mona_{t+1,k}-1 \right)
z^k \\
& +& z \sum_n {\tau_{t+1,n} \over {t
     \choose n}} \oint {dx \over x^{n+1}} \vev{ \prod_k \left( 
z^k \left( 1+x \right) ^k +1-z^k 
\right) ^{N_k(t)} } \nonumber
\end{eqnarray}

This equation is exact for finite $t$. To get the equation governing
the large $t$ behavior, we make a thermodynamic assumption : because
$T_t(z)$ converges to $T(z)$ in such a way that the averages and
variances converge as well, we expect that $\vev{ N_{k}(t+1) }
-\vev{ N_k(t) }$ converges to some deterministic quantity
$C_k$, so that $\vev{ N_{k}(t+1) } -\vev{ N_k(t) }-C_k$
may fluctuate but is $o(1)$ at large $t$. Then, by Cesaro
convergence, the same is true of $\vev{ N_k(t) }/t-C_k$.

Then the left-hand side of eq.(\ref{exact}) is $\sim \sum_k C_kz^k$. Using 
$\mona_{t+1,k} \simeq 1-{\moy  k \over t}$ for large $t$, the first
sum on the right-hand side is $\sim - \moy \sum_k kC_kz^k$.
In the second sum, we replace $x$ by $x/t$, leading to 
$$\sum_n {\tau_{t+1,n}t^n \over {t \choose n}} \oint {dx \over
  x^{n+1}} \vev{ \prod_k \left( z^k \left( 1+x/t \right) ^k +1-z^k
  \right) ^{N_k} }.$$
For large $t$, ${\tau_{t+1,n}t^n \over {t
\choose n}}\sim \tau_n$ and $\left( z^k \left( 1+x/t \right) 
^k +1-z^k \right) ^{N_k(t)}\sim \left(1+xkz^k/t\right)^{N_k(t)}
\sim e^{xkz^kN_k(t)/t}$. The averages are now trivial to take. The
second sum is seen to be
$$\sim \sum_n \tau_n \oint {dx \over
  x^{n+1}} e^{x\sum_k kz^kC_k}.$$
Putting this together, we get
$$\sum_k C_kz^k=-\moy \sum_k k C_kz^k+z\sum_n \tau_n \oint {dx \over
  x^{n+1}} e^{x\sum_k kz^kC_k}.$$
Defining $C(z)\equiv \sum_k C_kz^k$ and evaluating the contour
integral, we obtain the compact formula
\begin{equation}
C(z)=-z\moy \partial_z C(z)+z T\left( z \partial_z C(z) \right)
\label{evol_2}
\end{equation}
which describes the number of components of size $k$ at large $t$.

This equation has a simple interpretation in terms of destruction and
creation of components. Indeed, $C(z)=N_{t+1}(z)-N_t(z)$ gives for
each $k$ the variation of $N_k$ between $t$ and $t+1$. This variation
is the sum of two terms :
\begin{itemize}
\item $-z\moy \partial_z C(z)$ gives the product of the average number,
  $\moy $, of incoming edges in a given vertex by the mean proportion
  of vertices in components of size $k$ ($k C_k$). Stating that in the
  thermodynamic limit, two edges arriving at a vertex have vanishing
  probability of coming from the same finite component, this term
  can be seen as the average number of destroyed components.
\item $z T\left( z \partial_z C(z) \right)$ involves a sum of terms of
  the form ${\tau_n z^{1+k_1+\cdots +k_n} \over n!} k_1 C_{k_1}\cdots  k_n
  C_{k_n}$. It counts the average size of the component created when a
  new vertex is added.
\end{itemize}

Equation (\ref{evol_2}) allows a recursive calculation of the $C_k$'s.
Unfortunately, we were not able to find an explicit formula valid for
all $k$ and we only give here the values of the first few of them :
\begin{eqnarray*}
 C_1 & = & {\tau_0  \over 1+\moy } \\
 C_2 & = & {\tau_0 \tau_1  \over (1+\moy )(1+2\moy )} \\
 C_3 & = & {4\tau_0 \tau_1 ^2(\moy +1)+\tau_0 ^2\tau_2 (2\moy +1) 
\over 2(\moy +1)^2(2\moy +1)(3\moy +1)}
\end{eqnarray*}

\subsection{A simple example}
\label{simpleex}

The case when $T(z)=(1-p)+pz$ leads to tractable equations even at
finite $t$. As we shall see
later, this example is pathological from the point of view of the
percolation transition. This is another reason to give a separate treatment.

Though it is likely that this toy model has been solved more than once
before, we have found no reference. So we give a sketch of the
solution with apologies to the original contributions. In particular,
we compute the scaling function governing the statistics of large
components.

When vertex $t+1$ is added, it remains isolated with probability $1-p$
in which case $N_k(t+1)-N_k(t)=\delta_{k,1}$, or is attached to a
component of size $l$ to build a component of size $l+1$ with
probability $plN_l(t)/t$, in which case
$N_k(t+1)-N_k(t)=\delta_{k,l+1}-\delta_{k,l}$.  We infer
\begin{eqnarray*}\vev{N_{t+1}(z)}-\vev{N_t(z)} & = & (1-p)z
+\sum_{k,l}pl\vev{N_l(t)}/t(\delta_{k,l+1}-\delta_{k,l})z^k \\
& = & (1-p)z+p\sum_{l}l\vev{N_l(t)}/t(z^{l+1}-z^l) \\ & = &
(1-p)z+p/t(z^2-z)\partial_z\vev{N_t(z)}.\end{eqnarray*}
The initial condition is $\vev{N_1(z)}=N_1(z)=z$
We can simplify this equation by the change of variable $z=w/(1+w)$.
Setting $Q_t(w)\equiv \vev{N_t(z)}$, the equation for $Q_t$ is
$$Q_{t+1}-Q_t=(1-p)\frac{w}{1+w}-p\frac{w}{t}\partial_w Q_t.$$
$N_t(z)$ is a polynomial in $z$, so $Q_{t+1}$ has a regular series
expansion in $w$, $Q_{t} \equiv \sum_{k\geq 1} q_k(t)w^k$ which leads
to $q_k(t+1)=\frac{t-pk}{t}q_k(t)+(1-p)(-1)^{k+1}$. Direct
substitution shows that $q_k(t)=t\frac{1-p}{1+pk}(-1)^{k+1}$ is a
particular solution. The general solution $r_k(t)$ of the associated
homogeneous equation $r_k(t+1)=\frac{t-pk}{t}r_k(t)$ is 
$r_k(t)=\frac{\Gamma (t-pk) }{\Gamma (t)\Gamma (1-pk)}r_k(1)$. Taking
into account the initial condition $Q_{1}=w/(1+w)$ leads to 
$$q_k(t)=(-1)^{k+1}\left(t\frac{1-p}{1+pk}+\frac{p(k+1)}{1+pk}
\frac{\Gamma (t-pk) }{\Gamma (t)\Gamma (1-pk)}\right).$$
We can now go back to the $z$ variable :
$$\vev{N_k(t)}=\sum_{l=1}^{k} (-1)^{l+1}\left(t\frac{1-p}{1+pl}+
\frac{p(l+1)}{1+pl} \frac{\Gamma (t-pl) }{\Gamma (t)\Gamma (1-pl)}
\right)\frac{\Gamma (k) }{\Gamma (l)(k-l)!}.$$
The contribution corresponding to the first term inside the
parenthesis, the one which is linear in $t$, can be resummed. 
Indeed, if $C(z)$ is a solution of
$C=(1-p)z+p(z^2-z)\partial_zC$, $tC(z)$ solves the original equation
for $\vev{N_t(z)}$ even at finite $t$. The solution which is regular
at $z=0$ is $$C(z)=\sum_{k\geq 1} \frac{1-p}{p}\frac{\Gamma (k) 
\Gamma(1+1/p)}{\Gamma(k+1+1/p)}z^k.$$
One can check that, for any $p\in [0,1[$, $C'(z=1)=1$. As we shall
explain later, that means there is no percolation.

To summarize,
$$\vev{N_k(t)}=t\frac{1-p}{p}\frac{\Gamma (k) 
\Gamma(1+1/p)}{\Gamma(k+1+1/p)}+\sum_{l=1}^{k}(-1)^{l+1}
\frac{p(l+1)}{1+pl} \frac{\Gamma (t-pl) }{\Gamma (t)
\Gamma (1-pl)}\frac{\Gamma (k) }{\Gamma (l)(k-l)!}.$$

It is now possible to get the exact scaling function governing the
size distribution of large components. When both $t$ and $k$ are
large, the thermodynamic contribution scales like $
\frac{1-p}{p}\Gamma(1+1/p)tk^{-1-1/p}$ while in the other
contribution, for fixed $l$, one finds $(-1)^{l+1}\frac{p(l+1)}{1+pl}
\frac{1}{\Gamma (1-pl)\Gamma (l)}t^{-pl}k^{l-1}$. This decreases very
fast with $l$ so for large $k$ one can extend the range of $l$ to
$\infty$. The balance between
the thermodynamic and finite size contributions shows that the scaling
variable is $s=kt^{-p}$. In the scaling limit
$$\vev{kN_k(t)}\sim \frac{1-p}{p}\Gamma(1+1/p)s^{-1/p}+
\sum_{l\geq 1}(-1)^{l+1}\frac{p(l+1)}{1+pl}
\frac{1}{\Gamma (1-pl)\Gamma (l)}s^{l}.$$
Let $S(s)$ denote the scaling function on the right hand side.

Defining $sA(s)\equiv \sum_{l\geq 1}(-1)^{l+1}
\frac{1}{\Gamma (1-pl)\Gamma (l)}s^{l}$, one can rewrite
$$S(s)/s = 
\frac{1-p}{p}\Gamma(1+1/p)s^{-1-1/p}+A(s)-\frac{1-p}{p}\int_0^1
d\lambda  A(\lambda s)\lambda^{1/p}.$$

This expression of $S$ exhibits clearly its small $s$ behavior.  To
get control on the large $s$ behavior, one can use the familiar
representation $\frac{1}{\Gamma(z)}=\frac{1}{2i\pi}\int_{\mathcal C}
dw e^w w^{-z}$ for $z=1-pl$ and sum over $l$ to obtain
$$A(s)=\frac{1}{2i\pi}\int_{\mathcal C} dw e^{w-sw^p}w^{p-1}.$$
Observing that $\Gamma(1+1/p)s^{-1-1/p}=\int_0^\infty
d\lambda  A(\lambda s)\lambda^{1/p}$, one gets a compact expression
$$S(s)/s =A(s)+\frac{1-p}{p}\int_1^\infty
d\lambda  A(\lambda s)\lambda^{1/p}.$$

The large $s$ expansion can then be obtained by standard methods
(saddle point for A(s) and then analysis at $\lambda=1$ for the second
term). The formul\ae\ are rather cumbersome and we simply quote the
leading exponential behavior
$$ \log S(s) \sim -\frac{1-p}{p}p^{\frac{1}{1-p}}s^{\frac{1}{1-p}} 
\qquad s \to +\infty,$$ 
showing that $S(s)$ decreases very fast at large $s$.
\begin{figure}
\begin{center}

\includegraphics[width=.65\textwidth]{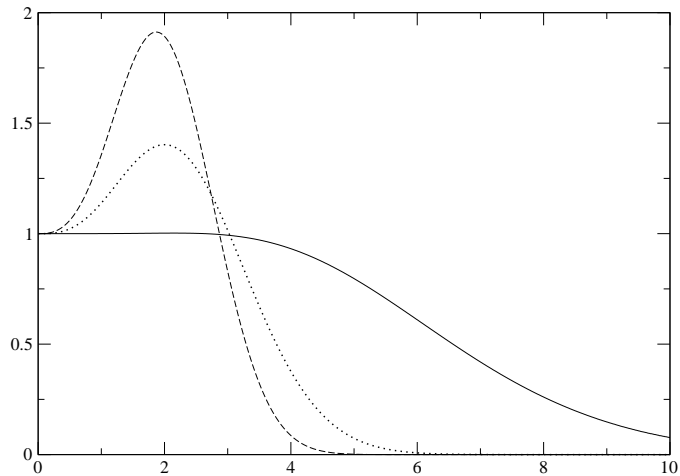}
\caption{Plots of $S(s)/S_{thermo}(s)$ for $p=0.2$ (solid line),
  $p=0.5$ (dots) and $p=0.6$ (dashed line)}
\end{center}
\end{figure}
As a simple example, take $p=1/2$. Then
$A(s)=\frac{1}{\sqrt{\pi}}e^{-s^2/4}$ and 
$$S(s)=\frac{1}{\sqrt{\pi}}\left(s e^{-s^2/4}+
\frac{1}{s^2}\int_s^\infty d\lambda
\lambda^2e^{-\lambda^2/4}\right).$$

Returning to arbitrary $p$, the relation $\vev{kN_k(t)}\sim S(k/t^p)$
implies that the average number of components of size $k \geq st^p$ is 
$\vev{\sum_{k \geq st^p} N_k(t)} \sim \int_s^\infty du \, S(u)/u
 < \infty$. So the large components have a size of order
$t^p$.

\subsection{Local-in-time results}
\label{local}

Knowing the global-in-time distribution of connected components, it is
natural to wonder whether this distribution is homogeneous in
time or not. More precisely, if $j=\sigma t$ with fixed $\sigma$ and
$t \rightarrow \infty$, what can we say about the probability for
vertex $j$ to be in a component of size k ?
\\
For $k \geq 1$ let $\rho_k(\sigma)$ be the probability that, in the
thermodynamic limit, vertex $\sigma t$ belongs to a connected component
of size $k$. The purpose of this section is to give an equation
governing the $\rho_k$'s.

\paragraph{Local-in-time equation}
The systematic to derive this equation relies on the same ideas as
those used in paragraph~\ref{global} to get the global formula, and
involves heavy formul\ae. Consequently we use directly the
thermodynamic assumption and give a more intuitive argument.
\\
Let $\Delta t$ be an interval of time such that $\Delta t \gg 1$ but
${\Delta t \over t} \ll 1$. Between times $t$ and $t+\Delta t$, the
number of vertices added to the graph is large but much smaller than
the size of the graph.
\\
Following the interpretation of eq.~(\ref{evol_2}), the number of new
components of size $k$ is
$$\Delta t \sum_m {\tau_m \over m!} \sum_{k_1,\cdots ,k_m} k_1
C_{k_1}\cdots  k_m C_{k_m} \ \delta_{1+\sum_i k_i,k}$$
Each of the $m$
old components has average time distribution ${\rho_{k_i}(\sigma)
  \over C_{k_i}}$ so that, on average, the contribution of the $\Delta
t$ new components to the time distribution of components on $k$
vertices is
$$\Delta t \sum_m {\tau_m \over m!} \sum_{k_1,\cdots ,k_m} k_1
C_{k_1}\cdots  k_m C_{k_m} \left({\rho_{k_1}(\sigma)d\sigma \over
    C_{k_1}}+\cdots +{\rho_{k_m}(\sigma)d\sigma \over C_{k_m}} \right) \ 
\delta_{1+\sum_i k_i,k} $$

Let us put $\rho(\sigma,z) \equiv \sum_k \rho_k(\sigma)
z^k$. Multiplying the expression above by $z^k$ and summing over $k$
yields the term $\Delta t z^2 \partial_z \rho(\sigma,z)
T'(z\partial_zC)d\sigma$.

Following again the interpretation of eq.~(\ref{evol_2}), $\moy \Delta t k
C_k$ components of size $k$ are destructed between $t$ and $t+\Delta
t$. According to the definition of $\rho_k(\sigma)$, the number of
vertices of relative age contained between $\sigma$ and
$\sigma+d\sigma$, which belonged to one of these components is $\moy
\Delta t k \rho_k(\sigma)$.

After summation over $k$, the destruction term is $\moy \Delta t z
\partial_z \rho(\sigma,z) d\sigma$.
\\
Moreover, a vertex of relative age $\sigma$ at time $t$ has relative
age ${\sigma t\over t+ \Delta t}$ at time $t+\Delta t$. Hence, the
local-in-time profile verifies, to first order in $ \Delta t/t$, the
following relation
$$td\sigma \left( \rho\left({\sigma t \over t+\Delta t},z
  \right)-\rho(\sigma,z) \right)=-\moy \Delta t z \partial_z
\rho(\sigma,z) d\sigma+\Delta t z^2 \partial_z \rho(\sigma,z)
T'(z\partial_zC)d\sigma.$$
This leads to the differential equation
\begin{equation}
\sigma \partial_\sigma \rho =\left( \moy -z T' \left( z
    \partial_z C \right) \right) z \partial_z \rho
\label{evol_3}
\end{equation}

As in the case of a Poisson law, this equation leads to recursion
relations for the $\rho_k(\sigma)$'s. Differentiating
eq. (\ref{evol_3}) $k$ times with respect to $z$ and taking $z=0$
leads to a first order linear differential equation for $\rho_k$, in
which $\rho_1,\cdots ,\rho_{k-1}$ appear. Putting $x=\sigma^{\moy }$, the
first few distributions are~:
\begin{eqnarray*}
\rho_1 & = & \tau_0 x\\
\rho_2 & = & {\tau_0 \tau_1  \over \moy } \left[ x-{x^2 \over 1+ \moy }
\right] \\
\rho_3 & = & x \left[ {\tau_0 ^2 \tau_2   \over
    2\moy (1+\moy )}+{\tau_1 ^2\tau_0  \over \moy ^2} \right] \\
& & - x^2 {2\tau_0 \tau_1 ^2 \over \moy ^2 (1+\moy )} \\
& & + x^3 \left[ {\tau_0 \tau_1 ^2 \over \moy ^2 (1+2\moy )}-{\tau_0 ^2\tau_2 
    \over 2\moy (1+\moy )^2} \right]
\end{eqnarray*}

\begin{figure}
\begin{center}
\includegraphics[angle=-90,width=.65\textwidth]{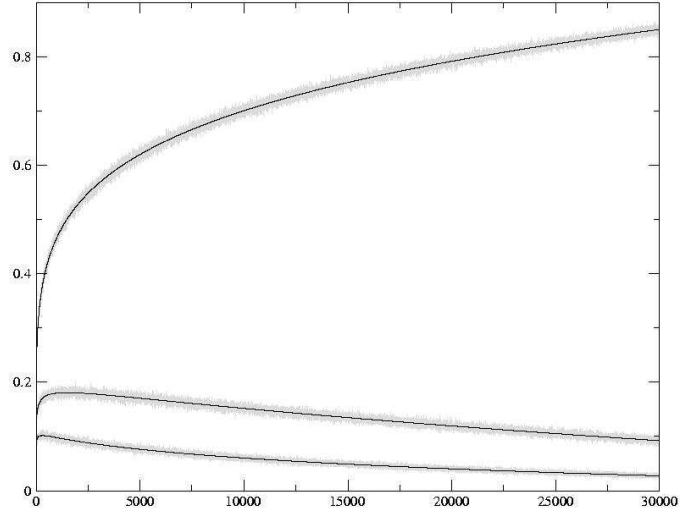}
\caption{The analytic result  (solid lines) for the profiles of small connected
  components (from top to bottom k=1,2,3) compared to numerical
  simulations (gray clouds) on 5000 random graphs of size
  30000. $T(z)=(1-p)/(1-pz)$ with $p=0.2$.}
\label{figprofils}
\end{center}
\end{figure}

\paragraph{Component distribution at $\sigma=1$}
To conclude this paragraph, we shall compute the probability for the
oldest vertex to belong to a component of size $k$. This is not
difficult to do because, unlike all the other vertices in the graph,
the youngest one does not have any outgoing edge. Hence, following
once again the interpretation of eq.~(\ref{evol_2}), it belongs to a
component of size $k$ with probability $\sum_m {\tau_m \over m!}
\sum_{k_1,\cdots ,k_m} k_1 C_{k_1}\cdots  k_m C_{k_m} \ \delta_{1+\sum_i
  k_i,k}$.  More compactly~:
\begin{equation}
\rho(\sigma=1,z)=zT(z\partial_zC).
\label{rhoplusjeune}
\end{equation}
Note that eqs.(\ref{evol_3},\ref{rhoplusjeune}) can be used to recover the global
equation (\ref{evol_2}).
\subsection{Tree distributions}
\label{trees}
Given a connected graph $G$, one may wonder how many connected
components of the random evolving graph are isomorphic to $G$. In
other words, if $k$ is the number of vertices of $G$, we look for the
average number of increasing maps $v : [1,\cdots ,k] \rightarrow
[1,\cdots ,t]$ such that the vertices $v_1,\cdots ,v_{k}$ span a connected
component of the random graph isomorphic to $G$.
\\
Let $m_i$ be the number of edges incoming to vertex $i$ in $G$. The
probability that vertices $v_1,\cdots ,v_{k}$ span a connected component
of the random graph isomorphic to $G$ is evaluated using the
following two rules :
\begin{itemize}
\item Vertex $v_i$ has $m_i$ incoming edges coming from the given
  vertices ;
\item The vertices $w$ of the random graph which are not in the image
  of $v$ must not be connected to any of the $v_i$'s.
\end{itemize}
Hence, putting $v_{k+1}=t+1$, the probability we look for is
$$\prod_{i=1}^{k} \left[ {\tau_{v_i,m_i} \over {v_i-1 \choose
      m_i}}\prod_{v_i < w_i < v_{i+1}} \sum_{j \geq 0} {\tau_{w_i,j}
    \over {w_i-1 \choose j}} {w_i-i-1 \choose j} \right]$$

The average number of components is then obtained by summing this
expression over all increasing maps $v$
$$\vev{ n_G } = \sum_{1 \leq v_1 <\cdots <v_{k} \leq t}
\prod_{i=1}^{k} \left[ {\tau_{v_i,m_i} \over {v_i-1 \choose m_i}}
  \prod_{v_i < w_i < v_{i+1}} \sum_{j \geq 0} {\tau_{w_i,j} \over
    {w_i-1 \choose j}} {w_i-i-1 \choose j} \right]$$

In the thermodynamic limit, the sum over $j$ is given by formula
(\ref{theta}), and its product over $w_i$ tends to $e^{-i \moy  \ln
  {v_{i+1} \over v_i}}$ provided only large $w_i$'s contribute
significantly. Hence, approximating the sum over $v_1,\cdots ,v_{k}$ by
an integral yields a contribution of $G$ equal to
\begin{equation}
C_{m_1,\cdots ,m_k}=t^{k-m} \left( \prod_{i=1}^{k}\tau_{m_i} \right) \int_{0 \leq
  \sigma_1 \leq \cdots  \leq \sigma_k \leq 1} d\sigma_1
\sigma_1^{\moy -m_1}\cdots  d\sigma_k \sigma_k^{\moy -m_k}
\label{arbres}
\end{equation}
This formula looks pretty much like the one proposed in the
poissonnian case in \cite{bb}. In particular, it shows exactly in the
same way that only connected graphs with $k=m+1$ (i.e. \textit{trees}
by Euler's formula) give a contribution that scales like $t$ in the
thermodynamic limit. Moreover, the contribution of a tree with
degree distribution $m_i$ is
\begin{equation}
\prod_{i=1}^k {\tau_{m_i} \over i (\moy  + 1) - (m_1 +\cdots +m_i)}
\label{arbres_2}
\end{equation}
This formula was obtained by integration over all relative ages
$\sigma_0,\cdots ,\sigma_{k-1}$. However, if we only integrate over some
of these variables while fixing the others, say
$\sigma_{k_1}<\sigma_{k_2}<\cdots <\sigma_{k_p}$, we expect to get the
contribution of a given tree with $k$ vertices amongst which $p$
vertices have imposed age. For instance, taking $k=2$ and integrating
over $\sigma_1$ while imposing $\sigma_2=\sigma$  gives the
contribution of trees of size 2 with younger vertex of age
$\sigma$. Explicit integration yields a contribution $\tau_0 \tau_1
\sigma^{2\moy} / (\moy+1)$ . On the other hand, if we integrate
over $\sigma_2$ and fix $\sigma_1=\sigma$, we get $\tau_0 \tau_1
(\sigma^\moy/\moy-\sigma^{2\moy}/\moy)$ for the contribution of
trees of size 2 with older vertex of age $\sigma$. The sum of these
two quantities, which is expected to be the probability that a site of
relative age $\sigma$ belongs to a tree with two vertices is, indeed,
equal to $\rho_2(\sigma)$ as calculated from equation (\ref{evol_3}).
\\
In fact, this result is not really surprising, since the derivation of
eq.(\ref{evol_2}) relies on the fact that, in the thermodynamic
limit, a new vertex connects with vanishing probability to several
vertices in the same component. In other words, this equation of
evolution takes only trees into account so eq.(\ref{arbres_2}) should
imply it. The full proof is instructive but tedious, and we relegate
it to the
appendix.

\section{The percolation transition}

Formula (\ref{evol_2}) gives a relation between the $C_k$'s, which
represent the asymptotic number of connected components of size $k$.
This equation involves in particular the function $z \partial_z
C(z)=\sum_k k C_k z^k$.
\\
Note that $k C_k$ is the fraction of sites belonging to components of
size $k$. This means that, if $\sum _k k C_k=1$, no single component
in the graph can have size $O(t)$. On the other hand, if $\sum _k k
C_k<1$, the possibility exists that a giant component contains a
finite fraction of the sites. The possibility that several very large
components coexist is usually ruled out because under a rearrangment
of a number $o(t)$ of edges these components would merge with finite
probability.  Though we have not tried to build a formal argument in
the case of evolving graphs, the intuition remains the same and is
confirmed by numerical simulations. So we take for granted that if
$\sum _k k C_k<1$ a single giant component contains a fraction $1-\sum
_k k C_k$ of the sites.

\subsection{Main results.}

The arguments that lead to the main qualitative and quantitative
features of the percolation transition are of technical nature. So we
postpone them to the next section.

\paragraph{Percolation criterion.} 
Unless $T(z)$ is a polynomial of degree 1, the system contains a giant
component if either the variance $\var$ of $T$ is $>1/4$ or the
average $\moy$ of $T$ is $>1/2$. The boundary separating the non
percolating phase from the percolating phase is given by the two
conditions $\var=1/4$ and $\moy < 1/2$.  The emergence of a giant
componant is purely due to an increase of the variance of $T$ above
the threshold $1/4$.

\paragraph{Behaviour below threshold.}

If  $\var \leq 1/4$ and $\moy \leq 1/2 $, the distribution of
component sizes in the system is critical, caracterized by the
following three equivalent properties. 

The dominant singularity of $C(z)$ at $z=1$ is 

\begin{equation}
\label{Csing}
C^{sing}(z) \propto
(1-z)^{\frac{2}{1-\sqrt{1-4\var}}}.
\end{equation}   

For large $k$, the fraction of sites belonging to components of size
$k$ decreases like 
\begin{equation}
\label{kdecrbelow}
kC_k \propto k^{-\frac{2}{1-\sqrt{1-4\var}}}.
\end{equation}

For a finite system of size $t \to \infty$, the large components have
a size of order 
\begin{equation}
\label{grosclu}
k(t)\propto t^{\frac{1-\sqrt{1-4\var}}{2}}.
\end{equation} 

\paragraph{Behaviour above but close to threshold.}

The percolation transition is of infinite order when $\var$ crosses
the value $1/4$ while keeping $\moy < 1/2$. If we denote by
$P_{\infty}$ the fraction of sites occupied by the giant component,
then when $\var-1/4 \to 0^+$, $P_{\infty}$ is exponentially small :
\begin{equation}
\label{sing}
  \log P_\infty \sim -\pi/\sqrt{4\var-1} \quad
  \mbox{for } \var \to 1/4^+.
\end{equation}

For large $k$, the fraction of sites belonging to components of size
$k$ decreases like 
\begin{equation}
\label{kdecrabove}
kC_k \propto k^{-\frac{3}{2}}e^{-kP_\infty}.
\end{equation}

\subsection{Discussion.}

We start our discussion by analysing the behavior of $C(z)$ close to
$z=1$. As before, we assume that $T(z)$ is not a polynomial of degree
$1$, i.e. that $T''(1)\neq 0$.

We start from equation (\ref{evol_2}) and apply the operator
$z\partial_z$ to get
\begin{equation}
\label{diffevol_2}
(z\partial_z C) +\moy z\partial_z (z\partial_z C)=zT(z\partial_z C)
+zT'(z\partial_z C)z\partial_z (z\partial_z C)\end{equation}
 which involves only $z\partial_z C$.

Set $z=e^{\tau}$ and $Y(\tau) \equiv 1-z\partial_z C$ .
Equation (\ref{diffevol_2}) can be rewritten for $Y$ :
$$0=Y(\tau)-1+\moy \dot{Y}(\tau)+e^\tau
\left(T(1-Y(\tau))-T'(1-Y(\tau))\dot{Y}(\tau)\right).$$

We know make two assumptions. 

i) there is no giant
component : $\sum_k kC_k=1$ or equivalently $Y(0)=0$, 

ii) the size distribution of clusters has a first moment :
$\mu_1=\sum_k k^2C_k$ is finite.  
\\
Note that ii) implies that $\dot{Y}(0^-)$ exists
and has value $-\mu_1$.

We differentiate the
equation for $Y$ with respect to $\tau$ and put $\tau=0$, yielding :
$$\mu_1^2 T''(1) +\mu_1 (2\moy -1)+1=0.$$
The discriminant of this
equation for $\mu_1$ is $1-4\var$ where $\var=T''(1)+\moy-\moy ^2$ is
the variance of the distribution $T$.
\\
The case when $T''(1)=0$, i.e. when $T(z)$ is affine, has some
pathologies, but it has already been treated in detail. 

If $T''(1)
>0$, the quadratic equation has two roots,

a) both real and positive if $1-2\moy $ and $1-4\var$ are positive,

b) both real and negative if $1-2\moy $ is negative but $1-4\var$ is
positive,

c) both complex if $1-4\var$ is negative.
\\
Clearly, only case a) is compatible with our two assumptions i) and
ii). In the sequel we shall take for granted that in this case, the
assumptions i) and ii) are indeed true. 

In cases b) and c), at least one of the assumptions must fail.
We show that it is i), the absence of giant component. To do that we
need a more precise analysis. 

Let us first give some properties of
$Y(\tau)$ for $\tau \leq 0$. By construction, $Y(\tau)$ has a convergent
expansion in powers of $e^{\tau}$ with negative coefficients (except
the first) and is bounded by $0$. So $Y$ is continuous decreasing on
$]-\infty,0]$. 

To obtain the large order behavior of $C_k$ we simplify
eq.(\ref{diffevol_2}) assuming that $\tau$ and $Y(\tau)$ are
small\footnote{Note that we do not assume that they are of the same
  order of magnitude.}. This is certainly a good approximation to
describe the small $\tau$ behavior of $Y(\tau)$ when there is no
percolation cluster because in that case $Y(\tau)$ is continuous and
vanishes at $\tau=0$.  It is also true close to a continuous phase
transition because $Y(0)$ is small.

Keeping only the dominant contributions yields
$$
\dot{Y}\left(T''(1)Y-\moy \tau\right)+(1-\moy )Y+\tau \simeq 0.$$
Setting $\var F=T''(1)Y- \moy \tau$, we derive the limit equation
\begin{equation}
\label{univ}
\var F\dot{F}+F+\tau=0.\end{equation}

This equation is the same as the one found and studied in
\cite{doro2,bb}. As the presentation in \cite{bb} is closer in spirit
to this one, this is the one we refer to in the sequel for details. In
the poissonian case, the authors showed carefully that eq.(\ref{univ})
indeed contains the quantitative universal features of the exact
cluster generating function. We take for granted that this remains
true for general $T$.

\hspace{.5cm}

When $\var < 1/4$, we write $\binf=\sqrt{1-4\var}$. 
The general integral of eq.(\ref{univ}) is
\begin{equation}
\label{below}
\left(1+\frac{1-\binf}{2}\frac{F}{\tau}\right)^{\frac{1+\binf}{2\binf}}
  \left(1 +\frac{1+\binf}{2}\frac{F}{\tau}\right)^{\frac{\binf-1}{2\binf}} =
    \frac{\mathrm{C}^{st}}{\tau}.\end{equation}
Suppose  $F(0)=0$, but $F(\tau)/\tau$ is not bounded close to
$\tau=0^-$. Then at a point where $F(\tau)/\tau$ is large,
eq.(\ref{below}) implies that $F(\tau)$ is of order one, a
contradiction. For analogous 
reasons, if $F(0)=0$ and $F(\tau)/\tau$ is bounded, 
$\lim_{\tau \rightarrow 0^-}1 +\frac{1+\binf}{2}\frac{F}{\tau}=0$ 
and then 
\begin{equation}
\label{expand}
F(\tau)+\frac{2}{1+\binf}\tau \propto
\tau^{\frac{1+\binf}{1-\binf}}\ll \tau.
\end{equation}

To summarize, if $\var < 1/4$ and $F(0)=0$, $F'(0)$ exists (and then 
$\mu_1=\sum_k k^2C_k$ is finite) so that i) implies ii): this means
that if $\var < 1/4$ but $\moy > 1/2$, i) has to be wrong, and $\sum_k kC_k <1$.

A word of caution is needed here. To get the limiting equation
(\ref{univ}), we have neglected terms of order $\tau^2$ in
(\ref{diffevol_2}). So strictly speaking, eq.(\ref{expand}) is correct
only if $\frac{1+\binf}{1-\binf}< 2$, i.e. $ 2/9 < \var <
1/4$. A more careful analysis, analogous to that sketched in \cite{bb}, would show that in general, in the
absence of a giant component, the small $\tau$ expansion of $F(\tau)$ 
starts with a standard taylor series in $\tau$ up to order
$\lfloor \frac{1+\binf}{1-\binf}\rfloor $ and then a leading singularity
proportional to  $\tau^{\frac{1+\binf}{1-\binf}}$.

\hspace{.5cm}

When $\var < 1/4$, we write $\bsup=\sqrt{4\var-1}$.  Eq.(\ref{univ})
implies that
\begin{equation}
\frac{1}{2}\log(\var  F^2 + \tau F+\tau^2) - \frac{1}{\bsup}
\arctan\left(\frac{\bsup F}{F+2 \tau}\right)
\label{above}.\end{equation}
is locally constant. As $F$ is continuous, the above quantity jumps by
$\pm \pi$ when $F+2 \tau$ changes sign.  To fix conventions, we
specify the function $\arctan$ by demanding that it is continuous and
takes value in $]-\pi/2,\pi/2[$. We argue by contradiction that $F$
cannot vanish at $\tau=0$ : if $F$ and $\tau$ are small, the argument
of the log is a small positive number so the first term gets large
and negative, while the term involving $\arctan$ remains bounded. So
again i) fails.

We have established the percolation criterion announced in the
previous section.

Furthermore, the relation $\var-1/4=T''(1)-(2\moy -1)^2/4$ shows that
the region separating the percolating phase from the non
percolating phase is $\var=1/4$ and $2\moy -1\leq 0$ : the emergence
of a giant component is purely due to an increase of the variance of
$T$ above the threshold $1/4$. This is a bit counterintuitive, because
it implies that there are cases with arbitrary small $\moy $ and a
giant component.

To illustrate that point, we present the case when $T$ is a quadratic
polynomial.  

\begin{figure}
\begin{center}
\includegraphics[width=.45\textwidth]{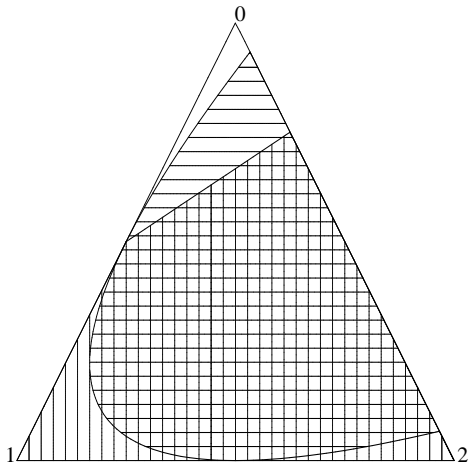}
\caption{Phase structure of the evolving graph in the subsimplex
  $0=p_3=p_4=\cdots $, $p_0+p_1+p_2=1$. The unshaded area corresponds to
  the non percolating phase. Horizontal lines fill the region $\gamma
  \geq 1/4$ and vertical lines the region $\moy \geq 1/2$.}
\label{percolquadra}
\end{center}
\end{figure}
The properties that describe the system when there is no giant
component or close to the threshold (when the giant component is
small) only rely on equation eq.(\ref{univ}), and we refer to
\cite{bb} for the detailed analysis that leads to
eqs.(\ref{Csing},\ref{kdecrbelow},\ref{grosclu},\ref{sing},\ref{kdecrabove}).

\subsection{Comparison with static graphs.}

Maximum entropy static graphs with fixed in- and/or out-degree
distributions have been studied in \cite{bb2,whittle}. If only the
in-degree is fixed, the out-degree is always Poissonian, in contrast
with the global geometric out-degree distribution of the evolving
cousin model. Let us compare briefly the percolation criteria of the
evolving case with two static maximum entropy situations. If
$T_-(z)=T(z)$ is the in-degree and $T_+(z)$ the out-degree
distribution of the static model, we read from 
\cite{bb2,whittle} that the condition for percolation is
$T''(1)T_+''(1)-(T'(1)-T'(1)^2)^2>0$ \footnote{This is, as should be
  the case for a static graph, symmetric in $T$ and
$T_+$ because $T'(1)=T_+'(1)$ is automatic : any edge is $in$ at one
end and $out$ at the other.} 
When only the in-degree is fixed in the static case, $T_+$ is
Poissonian, one finds $\var +\moy >1$. When moreover the out-degree is
fixed to be independent of the in-degree and follow a geometric law, 
so that the static graph has the same (global) degree distribution as
the evolving graph, one finds $2\var +\moy^2 >1$. The
thresholds are different in the static and evolving cases. Moreover,
the percolating region of the evolving case always contains strictly
the static percolating region, because when $\var <1/4$ and $\moy <
1/2$ both $\var +\moy$ and $2\var +\moy^2$ are less than $3/4$ : the
inhomogeneities of the evolving graph, in which old vertices have an
effective high coordinacy, favor the emergence of a giant
component. However, this giant component starts with a very tiny size,
in contrast with the static case, when its growth is generically linear.

\subsection{Comments on the profile of the infinite cluster in the
  percolating phase.}

 For a Poissonian in-degree distribution, the
authors of \cite{bb} obtained a closed equation which fitted
perfectly with numerical simulations. The naive
adaptation of their argument to the general case is straightforward,
but gives an incorrect result.

Though we have not been able to derive a closed equation for the
profile of the infinite component in the case of an arbitrary degree
distribution, the successive derivatives of this
profile at $\sigma=1$ can be computed in a systematic  way as follows.

Defining $D\equiv z\partial_zC$, we can derive from the previous
results that
\begin{eqnarray*}
\sigma \partial_\sigma \rho & = & \left( \moy -z  T' \left(D\right) \right) z \partial_z \rho\\
zT(D)-D & = & (\moy -zT'(D) )z \partial_z D \\
\rho(\sigma=1,z) & = & zT(D) .
\end{eqnarray*}
The first equation is just a rewriting of eq.(\ref{evol_3}), and the
second one was obtained by applying the operator $z\partial_z$ to eq.(\ref{evol_2}).The last equation gives the density at $\sigma=1$. Then the first
equation can be used to get the derivative of the density at
$\sigma=1$ :
\begin{eqnarray*}
\sigma \partial_\sigma \rho (\sigma=1,z) & = & \left( \moy -z T'
\left(D\right) \right) z \partial_z \rho(\sigma=1,z)\\
& = & \left( \moy -z T'
\left(D\right) \right)z \partial_z (zT(D)) \\
& = &\left( \moy -z  T'
\left(D\right) \right)(zT(D)+zT'(D)z\partial_z D) \\
& = & \left( \moy -z T' \left(D\right)
\right)zT(D)+zT'(D)(zT(D)-D) \\ 
& = & \moy zT(D)-zDT'(D).
\end{eqnarray*}
If  $D(z=1)=1$, $\rho(\sigma=1,z=1)=1$, but $\sigma \partial_\sigma
\rho (\sigma=1,z=1)=0$. 
The same kind of algebra can be used to compute the successive
derivatives $\left((\sigma
\partial_\sigma)^n \rho\right)(\sigma=1,z)$ for $n=2,\cdots $.
Again, one can check that this vanishes if $D(z=1)=1$. This is of
course natural in the non percolating phase. 

However, in the
percolating phase, $D(z=1)=1-P_\infty$ and one finds
\begin{eqnarray*}\rho(\sigma=1,z=1) & = & T(1-P_\infty)\\
\sigma \partial_\sigma \rho (\sigma=1,z=1) & = & 
\moy T(1-P_\infty)-(1-P_\infty)T'(1-P_\infty).
\end{eqnarray*}  

Formally, we can write
\begin{equation}
\rho (\sigma,z)=e^{\log \sigma [(\moy -z T'
  (D))z\partial_z+(zT(D)-D)\partial_D]}zT(D),
\label{develrho}
\end{equation}
where now $z$ and $D$
are independant variables, which is true order by order in an
expansion in powers of $\log \sigma$. Writing $\rho_\infty
(\sigma)=1-\rho(\sigma,z=1)$ for the profile of the giant component, eq.(\ref{develrho}) can be used to obtain
systematically, for small values of $n$, a (rather complicated)
formula for $\left((\sigma \partial_\sigma)^n
  \rho_\infty \right)(\sigma=1)$ as a polynomial in $P_\infty,
T(1-P_\infty),T'(1-P_\infty)\cdots  ,T^{(n)}(1-P_\infty)$.
\\
In particular, as a trivial example,
\begin{equation}
\rho_\infty (\sigma=1)=1-T(1-P_\infty) \simeq \moy P_\infty.
\label{rhoinf_1}
\end{equation}
On the other hand, for $\sigma$ close to $0$,
\begin{equation}
1-\rho_\infty (\sigma) \propto \sigma^\moy,
\label{rhoinf_0}
\end{equation}
so the giant component contains all the old vertices, but only a
fraction (which is exponentially small close the threshold) of the
young vertices. That means the percolation transition is very
inhomogeneous and takes mainly place in the part of the graph where it 
is denser than average.
\section{Conclusions.}

In this study, we have solved a model of evolving random graph which,
albeit simple, involves an arbitrary in-degree distribution. 

We have described the degree distributions and their local-in-time
profiles.  By construction, in-degree and out-degree at each vertex
are independent.  The local out-degree distribution follows an age
dependant Poisson law, which after integration over ages leads to a
geometric global out-degree distribution.

We have also made a detailed analysis of the distribution of component
sizes, again at the global and the local-in-time level. We have shown
the validity of the self averaging hypothesis by proving the
equivalence with a direct exact tree enumeration. 

The parameter
controlling the percolation transition has been found, quite
surprisingly, to be simply the variance of the in-degree
distribution, in contrast with the analogous static models. 

Below the transition, the large components have a size which scales like
a power of the total size of the graph. 
The size of the giant component close to the threshold
has been computed. It is exponentially small. 
The unusual fluctuation induced percolation mechanism might be the
reason why the critical behavior of this models is so different from
what is observed in the case of static graphs (see \cite{bb2,molloy,whittle}). There, the generic behavior close to the transition is
a linearly growing giant component. 

We have shown how all these differences could be used to discriminate
in certain cases between a static and an evolving random graph even
when the in and out degree distributions are the same for both. 

Among the unanswered questions is a direct description of the profile
of the giant component. For a Poissonian in-degree distribution, the
authors of \cite{bb} obtained a closed equation which fitted
perfectly with numerical simulations. The naive
adaptation of their argument to the general case is straightforward,
but gives an incorrect result. It would be desirable to find a valid
argument for general $T$, or more modestly to understand why the
argument in fact works for the Poissonian case.

\appendix

\section{A combinatorial identity.}

We show that the tree distribution, eqs.(\ref{arbres},\ref{arbres_2}), 
leads to the generating function formula, eq.(\ref{evol_2}).
\\
If $V$ is a finite (nonempty) set, we denote by ${\mathcal T}_{V}$ the set of trees with
vertex set $V$.

If moreover $V$ is totally ordered , we denote its supremum by $s_V$.
If moreover $|V|\geq 2$, we define ${\mathcal T}_{V}^n$ for
$n=1,2,\cdots ,|V|-1$ as the set of trees with vertex set $V$ such that
vertex $s_V$ has $n$ neighbors. Note that ${\mathcal T}_{V}^n$ is non
empty.

If $V=[1,\cdots ,k]$ (with the usual order) we denote ${\mathcal
  T}_{V}$ by ${\mathcal T}_{k}$ and if $k\geq 2$ we denote ${\mathcal
  T}_{V}^n$ by ${\mathcal T}_{k}^n$ for $n=1,\cdots ,k-1$. We shall
sometimes use the shorthand notation ${\mathcal
  T}_{k_1,\cdots ,k_n}\equiv {\mathcal T}_{k_1} \times \cdots  \times
{\mathcal T}_{k_n}$.

If $V$ is a totally ordered finite (nonempty) set, there is a unique
order preserving map from $V$ to $[1,\cdots ,|V|]$, so that there is a
canonical bijection between ${\mathcal T}_{V}$ and ${\mathcal
  T}_{|V|}$. For the same reason, if moreover $|V|\geq 2$ there is a
canonical bijection between ${\mathcal T}_{V}^n$ and ${\mathcal
  T}_{|V|}^n$ for any $n=1,2,\cdots ,|V|-1$.

\vspace{1cm}

Let $V$ be a totally ordered finite (nonempty) set $V$ with $|V|\geq
2$ and let $n\in [1,\cdots ,|V|-1]$. Take a tree $\mont \in {\mathcal
T}_{V}^n$. 

\vspace{.5cm}

To $\mont$ we associate the following data

\noindent $[i]\;\,\,$ An $n$-tuple of positive integers $(k_1,\cdots ,k_n)$ such
that $$k_1+\cdots +k_n=|V|-1.$$

\noindent $[ii]\;\,$ A sequence $(\mont_1,\cdots ,\mont_n) \in {\mathcal
  T}_{k_1,\cdots ,k_n}$.

\noindent $[iii]$ A sequence $(V_1,\cdots ,V_n)$ of disjoint subsets of $V$,
each endowed with the order induced from that of $V$, such that

$[a$] $(|V_1|,\cdots ,|V_n|)=(k_1,\cdots ,k_n)$,

$[b]$ $V_1 \cup \cdots  \cup V_n=V\backslash \{s_V\}$,  

$[c]$ $s_{V_1} < \cdots  < s_{V_n}$ for the order in $V$.

\noindent $[iv]\;$ A sequence $(v_1,\cdots ,v_n)\in [1,\cdots ,k_1]\times \cdots 
  \times [1,\cdots ,k_n]$,

\noindent as follows :

Remove from $\mont$ the vertex $s_V$ and the edges incident to it.
What remains is a forest made of $n$ components. There is a single way
to label the corresponding $n$ vertex sets $V_1,\cdots ,V_n$ so as to
satisfy $[c]$, and then we define $(k_1,\cdots ,k_n)$ by $[a]$ so we
have obtained $[i]$ and $[iii]$. For $l\in [1,\cdots ,n]$ the connected
component with vertex set $V_l$ is a tree.  We define $\mont_l$ as its
canonical representative in ${\mathcal T}_{k_l}$ and by $v_l$ the
vertex of $\mont_l$ whose preimage in $V_l$ is connected to $s_V$ in
$\mont$; this gives $[ii]$ and $[iv]$.

Conversely, one can recover $\mont$ from the data by reversing the
procedure.

We let ${\mathcal P}^V_{k_1,\cdots ,k_n}$ be the set of sequences
$(V_1,\cdots ,V_n)$ of disjoint subsets of $V$, each endowed with the
order induced from that of $V$, satisfying conditions $[a]$ and $[b]$
above. The set obtained when moreover $[c]$ is taken into account is
denoted by ${\mathcal P}^{restr\;V}_{k_1,\cdots ,k_n}$. There is a $1$
to $n!$ corespondence between ${\mathcal
  P}^{restr\;V}_{k_1,\cdots ,k_n}$ and ${\mathcal
  P}^{V}_{k_1,\cdots ,k_n}$

To summarize what we have found, we have put ${\mathcal T}_{V}^n$ in

-- $1$ to $1$ correspondence with the disjoint union $$\cup {\mathcal
  T}_{k_1,\cdots ,k_n} \times {\mathcal P}^{restr \;V}_{k_1,\cdots ,k_n}
\times [1,\cdots ,k_1]\times \cdots  \times [1,\cdots ,k_n],$$
  
-- $1$ to $n!$ correspondence with the disjoint union $$\cup
  {\mathcal T}_{k_1,\cdots ,k_n} \times {\mathcal
    P}^{V}_{k_1,\cdots ,k_n} \times [1,\cdots ,k_1]\times \cdots  \times
  [1,\cdots ,k_n],$$

\noindent where in both cases the union is taken over $n$-tuples of
positive integers $(k_1,\cdots ,k_n)$ such that $k_1+\cdots +k_n=|V|-1$.

\vspace{1cm}

From now on, we fix $V=[1,\cdots ,k]$ for some integer $k\geq 2$. We
assign to each tree $\mont \in {\mathcal T}_{k}$ a weight as follows.
Write $m_i$ for be the number of edges of $\mont$ connecting $i$ to
$[1,\cdots ,i-1]$. Then $${\mathcal W}_{\mont}= \tau_{m_1}\cdots 
\tau_{m_k} \int_{0\leq \sigma_1 \leq \cdots  \leq \sigma_k \leq 1}
d\sigma_1 \cdots  d\sigma_k \sigma_1^{\moy-m_1} \cdots 
\sigma_k^{\moy-m_k}.$$  Write ${\mathcal H}_{\mont}$ for the first
factor and ${\mathcal I}_{\mont}$ for the integral, so that ${\mathcal
  W}_{\mont}={\mathcal H}_{\mont}{\mathcal I}_{\mont}$.  We change
variables and set $\sigma_k=\tilde{\sigma_k}$ and
$\sigma_i=\tilde{\sigma}_i \tilde{\sigma}_k$ for $i=1,\cdots ,k-1$. The
power of $\tilde{\sigma}_k$ in the new integrand is $k-1+k\moy
-m_1-\cdots -m_k=k\moy$ so integration over $\tilde{\sigma}_k$ leads
to ${\mathcal I}_{\mont}=\frac{1}{1+k\moy}\tilde{\mathcal I}_{\mont}$
with $$\tilde{\mathcal I}_{\mont}= \int_{0\leq \sigma_1 \leq \cdots  \leq
  \sigma_{k-1} \leq 1} d\sigma_1 \cdots  d\sigma_{k-1}
\sigma_1^{\moy-m_1} \cdots  \sigma_{k-1}^{\moy-m_{k-1}}.$$
To avoid ambiguities when several trees are used at the same time, we
shall sometimes write $m^{\mont}_i$ instead of $m_i$.

Suppose that $\mont \in {\mathcal T}_{k}^n$, which amounts to set
$n=m_k$.  We want to express ${\mathcal W}_{\mont}$ in terms of its
decomposition, in fact in term of the $n$ trees
$(\mont_1,\cdots ,\mont_n)\in {\mathcal T}_{k_1} \times \cdots  \times
{\mathcal T}_{k_n}$ with $k_1+\cdots +k_n=|V|-1$, and of the partition
$V_1 \cup \cdots  \cup V_n$ (it turns out that ${\mathcal W}_{\mont}$
does not depend on the choice of one vertex in each $V_i$).

The decomposition procedure associates to each $i \in[1,\cdots ,k-1]$
one of the trees  $\mont_1,\cdots ,\mont_n$, say $\mont_l$ and a vertex
$i'\in [1,\cdots ,k_l]$ in $\mont_l$. By construction of the decomposition,
if  $j \in [1,\cdots ,i-1]$ is such that $(i,j)$ is an edge of $\mont$
then $i$ and $j$ have the same $\mont_l$, $j' \in [1,\cdots ,i'-1]$
and $(i',j')$ is an edge of $\mont_l$. Hence
$m^{\mont}_i=m_{i'}^{\mont_l}$ and ${\mathcal
H}_{\mont}=\tau_n {\mathcal H}_{\mont_1}\cdots  {\mathcal H}_{\mont_n}$
has a simple multiplicative behavior. 

Our aim is now to show that when $(\mont_1,\cdots ,\mont_n)$ (and then
automatically $(k_1,\cdots ,k_n)$) are fixed $\sum_{{\mathcal
 P}^{[1,\cdots ,k] }_{k_1,\cdots ,k_n}}\tilde{\mathcal
I}_{\mont}={\mathcal I}_{\mont_1} \cdots  {\mathcal I}_{\mont_n}$.

We introduce another tree, $\dot{\mont}$, whose decomposition is made
of the same trees $(\mont_1,\cdots ,\mont_n)$ as $\mont$, but with
$\dot{V}_1=[1,\cdots ,k_1],\dot{V}_2=[k_1+1,\cdots ,k_1+k_2],\cdots $ and
$\dot{v}_1=k_1,\dot{v}_2=k_1+k_2,\cdots $. Write $\dot{m}_i$ for the
number of edges of $\dot{\mont}$ connecting $i$ to $[1,\cdots ,i-1]$.

There is a unique permutation, say $\lambda$, of $[1,\cdots ,k-1]$
which maps $V_1$ into $\dot{V}_1$, $V_2$ into $\dot{V}_2$, $\cdots $,
and is strictly increasing in each.  Then $m_i=\dot{m}_{\lambda(i)}$
for each $i \in [1,\cdots ,k-1]$. If we set
$\sigma_i=\dot{\sigma}_{\lambda(i)}$ we obtain $$\tilde{\mathcal
I}_{\mont}=\int_{0\leq \dot{\sigma}_{\lambda(1)} \leq \cdots  \leq
\dot{\sigma}_{\lambda(k-1)} \leq 1} d\dot{\sigma}_1 \cdots 
d\dot{\sigma}_{k-1} \dot{\sigma}_1^{\moy-m_1} \cdots 
\dot{\sigma}_{k-1}^{\moy-m_{k-1}}.$$ Write $R_{\lambda}$ for the
region of integration $0\leq \dot{\sigma}_{\lambda(1)} \leq \cdots 
\leq \dot{\sigma}_{\lambda(k-1)} \leq 1$. 

Conversely, $(k_1,\cdots ,k_n)$ being kept fixed, if $(V_1,
\cdots , V_n)$ describes ${\mathcal
    P}^{[1,\cdots ,k]}_{k_1,\cdots ,k_n}$  each permutation $\lambda$
of $[1,\cdots ,k-1]$ such that $\lambda^{-1}$ is strictly increasing
when restricted to $\dot{V}_1,\cdots ,\dot{V}_n$ appears exactly once.
If $R$ be the union of all such $R_{\lambda}$'s (the intersection of
different $R_{\lambda}$'s is of measure 0), one checks that
$(\dot{\sigma}_1,\cdots ,\dot{\sigma}_{k-1})$ is in $R$ if and only if
$0 \leq \dot{\sigma}_1\leq \cdots  \leq \dot{\sigma}_{k_1}\leq 1$,
$0 \leq \dot{\sigma}_{k_1+1} \leq \cdots  \leq \dot{\sigma}_{k_1+k_2}\leq 1$,
$\cdots $.

This shows that when $(\mont_1,\cdots ,\mont_n)$ are fixed $\sum_{{\mathcal
P}^{[1,\cdots ,k] }_{k_1,\cdots ,k_n}}\tilde{\mathcal
I}_{\mont}={\mathcal I}_{\mont_1} \cdots   {\mathcal I}_{\mont_n}$.

If we recall moreover that ${\mathcal H}_{\mont}=\tau_n {\mathcal
  H}_{\mont_1}\cdots  {\mathcal H}_{\mont_n}$ and that ${\mathcal
  T}_{V}^n$ is in $1$ to $n!$ correspondence with the disjoint union
$$\cup_{(k_1,\cdots ,k_n)} {\mathcal T}_{k_1,\cdots ,k_n} \times
{\mathcal P}^{V}_{k_1,\cdots ,k_n} \times [1,\cdots ,k_1]\times \cdots 
\times [1,\cdots ,k_n],$$ we obtain as an immediate consequence that
$$(1+k\moy) \sum_{\mont \in {\mathcal T}_{k}^n} {\mathcal
  W}_{\mont}=\sum_{(k_1,\cdots ,k_n)} \sum_{(\mont_1,\cdots ,\mont_n) \in
  {\mathcal T}_{k_1,\cdots ,k_n}} \frac{\mont_n}{n!} k_1{\mathcal
  W}_{\mont_1}\cdots  k_n{\mathcal W}_{\mont_n}.$$

If we define ${\mathcal
  W}^{(k)} \equiv \sum_{\mont \in {\mathcal T}_{k}} {\mathcal
  W}_{\mont}=\sum_n \sum_{\mont \in {\mathcal T}_{k}^n} {\mathcal
  W}_{\mont}$, it is plain that $\sum_k {\mathcal
  W}^{(k)}z^k$ satisfies eq.(\ref{evol_2}), which has only one formal
power series solution vanishing at $z=0$. 
QED.


\begin{thebibliography}{}
  
\bibitem{barab} 
 R. Albert and A.-L. Barab\'asi, \textit{Statistical mechanics of
   complex networks}, Reviews of Modern Physics 74, 47 (2002). 
  
\bibitem{bb} M. Bauer and D. Bernard, \textit{A simple asymmetric
    evolving random network}, {\tt ArXiv:cond-mat/0203232}


\bibitem{bb2} M. Bauer and D. Bernard, \textit{Maximal entropy random
    networks with given degree distribution}, {\tt ArXiv:cond-mat/0206150}

\bibitem{new} 
   D. S. Callaway, J. E. Hopcroft, J. M. Kleinberg,
   M. E. J. Newman and H. Strogatz, 
   \textit{Are randomly grow graphs really random ?},
   Phys. Rev. E, Vol 64 (2001) 041902.


\bibitem{doro1} S. N. Dorogovtsev and J.F.F. Mendes, \textit{Evolution of
  Networks}, Adv. Phys. \textbf{51}, 1079, (2002).


\bibitem{doro2}
   S.~N.~Dorogovtsev, J.~F.~F.~Mendes,  and A.~N.~Samukhin,
\textit{Anomalous percolation properties of growing networks}
   Phys.\ Rev.\ E {\bf 64}, 066110 (2001).


\bibitem{erdos} P. Erd\"os and A. R\'enyi, \textit{On the evolution of
    random graphs}, Publ. Math. Inst. Hungar. Acad. Sci. 5 (1960),
  17--61.

\bibitem{kepes} N. Guelzim, S. Bottani, P. Bourgine and F. K\'ep\`es,
  \textit{Topological and causal structure of the yeast genetic
    network}, Nature Genetics, in press.

\bibitem{kim} J. Kim, P. L. Krapivsky, B. Kahng and S. Redner, 
   \textit{Infinite-order percolation and giant fluctuations in a
     protein interaction network}, Phys.\ Rev.\ E {\bf 66} (2002) 055101.

\bibitem{lee} Lee et al., \textit{Transcriptional regulatory networks
    in Saccharomyces Cerevisiae}, (2002) Science 298:799-804, see also 
{\tt http://web.wi.mit.edu/young/regulator\_network}

\bibitem{molloy} M. Molloy and B. Reed, \textit{Statistical mechanics
    of complex networks}, Random Struct. Algorithms
  \textbf{6}, 161, (1995); and Comb. Proba. Comput. \textbf{7}, 295, (1998). 

\bibitem{newm}  M. E. Newman, S. H. Strogatz and D. Watts,
  \textit{Random graphs with arbitrary degree distributions and their
    applications}, Phys. Rev. E
\textbf{64} (2001) 026118. 

\bibitem{whittle} P. Whittle, \textit{The statistics of random
    directed graphs}, J. Stat. Phys. \textbf{56}, 499, (1989).

\end{thebibliography}
\end{document}